\documentstyle[prd,aps,preprint]{revtex}

\draft

\input{epsf.sty}

\begin{document}


\title{The appearance of coordinate shocks in \\ hyperbolic formalisms
of General Relativity}

\author{Miguel Alcubierre\thanks{Present address: Max Planck Institut
f\"{u}r Gravitationsphysik, Albert Einstein Institut, Schlaatzweg~1,
D-14473~Potsdam, Germany.}}

\address{Department of Physics and Astronomy, University of Wales, \\
College of Cardiff, P.O. Box 913, Cardiff CF1 3YB, UK.}

\maketitle


\begin{abstract}{I consider the appearance of shocks in hyperbolic
formalisms of General Relativity.  I study the particular case of the
Bona-Mass\'{o} formalism with zero shift vector and show how shocks
associated with two families of characteristic fields can develop.
These shocks do not represent discontinuities in the geometry of
spacetime, but rather regions where the coordinate system becomes
pathological.  For this reason I call them `coordinate shocks'.  I show
how one family of shocks can be eliminated by restricting the
Bona-Mass\'{o} slicing condition \mbox{\,$\partial_t \alpha = -
\alpha^2 f(\alpha) \, {\rm tr} K$\,} to the case \mbox{\,$f = 1 +
k/\alpha^2$}, with \,$k$\, an arbitrary constant.  The other family of
shocks can not be eliminated even in the case of harmonic slicing
\mbox{\,$(f = 1)$}.  I also show the results of the numerical evolution
of non-trivial initial slices in the special cases of a flat two-dimensional
spacetime, a flat four-dimensional spacetime with a spherically symmetric
slicing, and a spherically symmetric black hole spacetime.
In all three cases coordinate shocks readily develop, confirming the
predictions of the mathematical analysis.  Although I concentrate in
the Bona-Mass\'{o} formalism, the phenomena of coordinate shocks should
arise in any other hyperbolic formalism.  In particular, since the
appearance of the shocks is determined by the choice of gauge, the
results presented here imply that in {\em any formalism}\/ the use of a
harmonic slicing can generate shocks.}\end{abstract}



\section{INTRODUCTION}

In the last few years there has been a renewed interest in the study of
initial-value formulations of General Relativity
\cite{BonaMasso89,BonaMasso92,BonaMasso95,BonaMasso96,ChoquetRuggieri83,%
ChoquetYork95a,ChoquetYork95b,Friedrich96,FrittelliReula96}.  This interest
has been motivated mainly by the desire of rewriting the Einstein system of
evolution equations in an explicit hyperbolic form, so that it can be
solved numerically using modern high resolution methods from fluid
dynamics~\cite{Leveque94}.

One can separate the new hyperbolic formalisms according to the way in
which they treat the evolution of the lapse function \,$\alpha$.
Some formulations assume the existence of an arbitrarily prescribed gauge,
{\em i.e.} the lapse is an arbitrary function of spacetime
known {\em a priori}\/~\cite{Friedrich96,FrittelliReula96} (`prescribed
gauge' formalisms).  Other formulations include the lapse function as
part of the system of dynamical variables, and postulate for it
an evolution equation that guarantees the hyperbolicity of the {\em whole}\/
system, geometry plus gauge
\cite{BonaMasso89,BonaMasso92,BonaMasso95,BonaMasso96,%
ChoquetRuggieri83,ChoquetYork95a,ChoquetYork95b} (`hyperbolic
gauge' formalisms).  The resulting formalisms remain hyperbolic for
any {\em prescribed}\/ shift vector.~\footnote{To date, in all these
hyperbolic formalisms the shift vector is assumed to be known {\em a priori}.
The author is aware, however, of some efforts to find evolution equations
for the shift that will keep the whole system hyperbolic~\cite{BonaMasso97}.}

Prescribed gauge formalisms, though certainly useful theoretically, might
have a limited applicability in Numerical Relativity simply because
there is no recipe that can give us the {\em a priori}\/ form of the
lapse except in trivial cases (for example \mbox{\,$\alpha =
1$}).  Hyperbolic gauge formalisms on the
other hand, by allowing the lapse functon to adapt itself to the
evolution of the geometry while maintaining the hyperbolic structure of
the system of equations, would appear to be much more promising.

Hyperbolic gauge formalisms, however, are probably more susceptible to
a problem that seems to have been overlooked until know.  By rewriting
the whole evolution system (gauge plus geometry) in hyperbolic form,
they open up the possibility of running into a well known non-linear
effect associated with hyperbolic systems: the appearance of shocks.
Here I use the term `shock' in a somewhat loose form to refer to a
discontinuous solution that develops from smooth initial data, without
worrying about the existence of weak solutions or jump conditions.

The fact that in vacuum General Relativity one can have shock fronts is
well \linebreak known
\cite{MTW,Pirani57,PapapetrouTreder62,Choquet68}.  By shocks fronts,
however, one generally understands discontinuities in the curvature of
spacetime present in the initial data that propagate with the speed of
light.  In the theory of non-linear hyperbolic equations such solutions
are not considered proper shocks, but are called instead `contact
discontinuities'.  Here, however, I will consider the existence of
discontinuous solutions that arise from smooth initial data even in a
flat spacetime.  Clearly those solutions do not correspond to a
physical discontinuity in the geometry of spacetime.  Instead the
discontinuities indicate regions where our coordinate system becomes
pathological: the time slices can become non-smooth, or a spatial
coordinate might map a finite proper distance to an infinitesimal
interval.  It is for this reason that I shall refer to them as
`coordinate shocks'.

Even though modern high resolution numerical methods can deal with the
presence of shock waves, clearly the appearance of coordinate shocks is
something that must be avoided.  In the first place, coordinate shocks
create completely artificial discontinuities in solutions that
otherwise represent perfectly smooth geometries. Not only that, but
since in general our gauge conditions are not obtained from a
conservation law, we will not have an analogue of the `weak solutions'
to such laws.  This means that after a shock forms our gauge conditions will
just break down, and even if the numerical solution remains well
behaved, {\em it will not have any clear physical meaning}.
In particular, as the numerical mesh is refined, the solution will not
converge after the formation of the shock.

\vspace{5mm}

In this paper I will concentrate in one particular hyperbolic formalism
of General Relativity, the Bona-Mass\'{o} (BM) formalism
\cite{BonaMasso89,BonaMasso92,BonaMasso95,BonaMasso96}, and I will show
how these coordinate shocks can and do indeed develop even in very
simple situations.

\section{THE BONA-MASS\'{O} FORMALISM}

In this section I will make a brief introduction to the BM hyperbolic
formalism for General Relativity.  I will use the most recent form of
this formalism as presented in~\cite{BonaMasso96}.

Let us start from the standard 3+1 formulation of General Relativity of
Arnowitt, Deser and Misner (ADM)~\cite{ADM62,York79}.  The evolution
equations for the metric \,$g_{ij}$\, and extrinsic curvature
\,$K_{ij}$\, are:  \begin{mathletters} \label{eq:ADM1} \begin{eqnarray}
\left( \partial_t \,-\, \cal{L}_\beta \right) \; g_{ij} &=& - 2 \,
\alpha \, K_{ij} \;\;, \\ \left( \partial_t \,-\, \cal{L}_\beta \right)
K_{ij} &=& - \nabla_i \, \nabla_j \, \alpha \,+\, \alpha \, \left[
R_{ij}^{(3)} \,+\, {\rm tr} K \, K_{ij} \,-\, 2 \, K_{ik} \, K^k_j
\,-\, R_{ij}^{(4)} \right] \;\;, \end{eqnarray} \end{mathletters}

\noindent where \,$\alpha$\, is the lapse function, \,$\beta^k$\, the
shift vector, and where \,$R_{ij}^{(3)}$\, and \,$R_{ij}^{(4)}$\,
represent the components of the Ricci tensor for the spatial
hyper-surfaces and for the full spacetime respectively.  In what follows
I will restrict myself to the case of zero shift vector.  The ADM
equations then reduce to:  \begin{mathletters} \label{eq:ADM2}
\begin{eqnarray} \partial_t \; g_{ij} &=& - 2 \, \alpha \, K_{ij} \;\;
,\\ \partial_t  K_{ij} &=& - \nabla_i \, \nabla_j \, \alpha \,+\,
\alpha \, \left[ R_{ij}^{(3)} \,+\, {\rm tr} K \, K_{ij} \,-\, 2 \,
K_{ik} \, K^k_j \,-\, R_{ij}^{(4)} \right] \;\; .  \end{eqnarray}
\end{mathletters}

In order to obtain a system that is first order in space we introduce
the following quantities:  \begin{equation} A_k \,=\, \partial_k \, \ln
\alpha \;\; , \qquad D_{kij} \,=\, {\displaystyle \frac{1}{2}} \;
\partial_k \, g_{ij} \;\;.  \end{equation}

The evolution equation for \,$K_{ij}$\, can then be rewritten as:
\begin{equation} \partial_t \, K_{ij} \,+\, \partial_k \, \left( \alpha
\, {\lambda^k}_{ij} \right) \,=\,  \alpha \, S_{ij} \;\; ,
\label{eq:Kev} \end{equation}

\noindent where we have defined:  \begin{equation} {\lambda^k}_{ij}
\,=\, {D^k}_{ij} \,+\, {\displaystyle \frac{1}{2}} \; \delta^k_i \left(
A_j \,+\, 2 \, V_j \,-\, {D_{jm}}^m \right) \,+\, {\displaystyle
\frac{1}{2}} \; \delta^k_j \left( A_i \,+\, 2 \, V_i \,-\, {D_{im}}^m
\right) \;\;, \label{eq:lambdadef} \end{equation}

\noindent with:  \begin{equation} V_k \,\equiv\, {D_{km}}^m \,-\,
{D^m}_{mk} \;\; .  \label{eq:Vdef} \end{equation}

The source term \,$S_{ij}$\, in equation~(\ref{eq:Kev}) involves only
the fields themselves and not their derivatives:  \begin{eqnarray}
S_{ij} &=& -\, R_{ij}^{(4)} +\, {\rm tr} K \, K_{ij} \,-\, 2 \, K_{ik}
\, K^k_j \,+\, 4 \, D_{kmi} \, {D^{km}}_j \,+\, {\Gamma^k}_{km}
{\Gamma^m}_{ij} \,-\, \Gamma_{ikm} {\Gamma_j}^{km} \nonumber \\ &&+
\left( A^k - 2 {D_m}^{km} \right) \left( \rule{0mm}{4mm} D_{ijk} +
D_{jik} \right) + A_i \left( V_j - {\textstyle \frac{1}{2}} \,
{D_{jk}}^k \right) + A_j \left( V_i - {\textstyle \frac{1}{2}} \,
{D_{ik}}^k \right) \;\;.  \end{eqnarray}

We also need an evolution equation for the \,$D_{kij}$.  This we obtain
by taking the spatial derivative of the evolution equation for
\,$g_{ij}$:  \begin{equation} \partial_t \, D_{kij} \,+\, \partial_k
\left( \rule{0mm}{4mm} \alpha \, K_{ij} \right) \,=\, 0 \;\; .
\label{eq:Dev} \end{equation}

The quantities \,$V_k$\, defined in~(\ref{eq:Vdef}) are very
important.  Their evolution equation can be obtained
from~(\ref{eq:Dev}).  In order to ensure hyperbolicity, however, it is
crucial to modify the resulting equation using the momentum constraints
to obtain:  \begin{equation} \partial_t \, V_k \,=\, \alpha \, P_k \;\;
, \label{eq:Vev} \end{equation}

\noindent where:  \begin{eqnarray} P_k &=& \rule{0mm}{5mm} \, G^0_k
\,+\, A_m \, \left( \rule{0mm}{4mm} K^m_k \,-\, \delta^m_k \, {\rm tr}
K \right) \,+\, K^m_n \left( \rule{0mm}{4mm} {D_{km}}^n \,-\,
\delta^n_k \, {D_{ma}}^a \right) \nonumber \\ && \rule{0mm}{5mm} -\, 2
\, K_{mn} \, \left( \rule{0mm}{4mm} {D^{mn}}_k \,-\, \delta^n_k \,
{D_a}^{am} \right) \;\; , \end{eqnarray}

\noindent and where \,$G_{\mu \nu}$\, is the Einstein tensor of the
spacetime.

The quantities \,$V_k$\, are now considered independent, and
equation~(\ref{eq:Vdef}) becomes an algebraic constraint that must be
satisfied by the physical solutions.

Finally, we need evolution equations for the lapse \,$\alpha$\, and its
derivative \,$A_k$\,, {\em i.e.} we need to choose a slicing
condition.  In the BM formalism the following slicing condition is
used:  \begin{equation} \partial_t \, \alpha \,=\, -\, \alpha^2 \, f
\left( \alpha \right) \, {\rm tr} K \;\; , \label{eq:alphaev}
\end{equation}

\noindent with \,$f(\alpha) \,>\, 0$\, but otherwise arbitrary.

The complete system of evolution equations then takes the form:
\begin{mathletters} \label{eq:evolution1} \begin{eqnarray} \partial_t
\, \alpha &=& -\, \alpha^2 \, f \left( \alpha \right) \, {\rm tr} K
\;\;, \\ \partial_t \; g_{ij} &=& -\, 2 \, \alpha \, K_{ij} \;\;,
\end{eqnarray} \end{mathletters}

\noindent and: \begin{mathletters} \label{eq:evolution2}
\begin{eqnarray} \partial_t \, A_k +\, \partial_k \left(
\rule{0mm}{4mm} \alpha f \, {\rm tr} K \right) &=& 0 \;\;, \\
\partial_t \, D_{kij} \,+\, \partial_k \left( \rule{0mm}{4mm} \alpha \,
K_{ij} \right) &=& 0 \;\;, \\ \partial_t \, K_{ij} \,+\, \partial_k \,
\left( \alpha \, {\lambda^k}_{ij} \right) &=&  \alpha \, S_{ij} \;\;,
\\ \partial_t \, V_k &=& \alpha \, P_k \;\;. \end{eqnarray}
\end{mathletters}

To study the characteristic structure of the system of
equations~(\ref{eq:evolution1}) and~(\ref{eq:evolution2}) we choose a
fixed space direction \,$x$\, and consider only derivatives along that
direction.  It can then be shown that the system is hyperbolic with the
following structure:~\footnote{Here I use the term hyperbolic in the
weak sense to mean that the characteristic matrix of the system has
real eigenvalues.  It should be noticed that this weak form of
hyperbolicity does not guarantee that the system can be diagonalised.
A crucial feature of the BM formalism is that even though it is only
weakly hyperbolic, it can in fact be diagonalised as long as \,$f >
0$\,.}

\begin{itemize}

\item 25 fields propagate along the time lines (zero speed).  These
fields are:  \begin{equation} \left\{ \rule{0mm}{4mm} \alpha,\,
g_{ij},\, A_{x'},\, D_{x'ij},\, V_i,\, A_x  - f {{D_x}^m}_m \right\}
\qquad \left( x' \,\neq\, x  \right) \;\; .  \end{equation}

\item 10 fields propagate along the physical light cones with speeds:
\begin{equation} {\lambda^l}_\pm \,=\, \pm \, \alpha \,
\sqrt{\rule{0mm}{3mm} g^{xx}} \;\; .  \end{equation}

These fields are:  \begin{equation} {w^l}_{ix'\pm} \,=\, K_{ix'}
\,\pm\, \sqrt{\rule{0mm}{3mm} g^{xx}} \, \left( \rule{0mm}{4mm}
D_{xix'}  \,+\, \delta^x_i \, V_{x'} / g^{xx} \right) \qquad \left( x'
\,\neq\, x  \right) \;\; . \label{eq:wl} \end{equation}

\item 2 fields propagate with the `gauge speeds':  \begin{equation}
{\lambda^f}_\pm \,=\, \pm \, \alpha \, \sqrt{f \, g^{xx}} \;\; .
\end{equation}

They are:  \begin{equation} {w^f}_\pm \,=\, \sqrt{f} \; {\rm tr} K
\,\pm\, \sqrt{\rule{0mm}{3mm} g^{xx}} \, \left( \rule{0mm}{4mm} A_x
\,+\, 2 \, V^x / g^{xx} \right) \;\; . \label{eq:wf} \end{equation}

\end{itemize}

\section{MATHEMATICAL ANALYSIS OF THE NON-LINEARITIES}

Here I will try to understand the nonlinearities present in our system
of equations, trying in particular to determine whether shocks can
develop.  From the discussion in the previous section it is clear that
the system of evolution equations in the BM formalism has the following
structure:  \begin{mathletters} \label{eq:system} \begin{eqnarray} &
\partial_t \, u_i \,=\, p_i & \qquad i \in \{1,...,N_u\} \;\;, \\ &
\partial_t \, v_i \,+\, \partial_x \, F_i \,=\, q_i & \qquad i \in
\{1,...,N_v\} \;\;.  \end{eqnarray} \end{mathletters}

The fluxes \,$F_i$\, that appear in the above equations have the form:
\begin{equation} F_i \,=\, \sum_{j=1}^{N_v} \; M_{ij} \, v_j \;\; ,
\end{equation}

\noindent where the coefficients \,$M_{ij}$\, are functions of the
\,$u$'s\, but not of the \,$v$'s.

Let us now call \,$\lambda_i$\, the eigenvalues and \,${\bf e}_i$\, the
corresponding eigenvectors of the Jacobian matrix \,$M_{ij} = \partial
\, F_i \,/\, \partial \, v_j$\,.  Let us also introduce the matrix \,$R
= [{\bf e}_1|{\bf e}_2|\cdots|{\bf e}_{N_v}]$\, of column
eigenvectors.  The eigenfields \,$w_i$\, are then defined by:
\begin{equation} {\bf v} \,=\, R \; {\bf w} \quad \Rightarrow \quad
{\bf w} \,=\, R^{-1} \, {\bf v} \;\;.  \end{equation}

A given eigenfield \,$w_i$ is called `linearly degenerate'~\cite{Lax72}
if the following condition holds:  \begin{equation} \frac{\partial \,
\lambda_i}{\partial \, w_i} \,=\, \sum_{j=1}^{N_v} \; \frac{\partial \,
\lambda_i}{\partial \, v_j} \, \frac{\partial \, v_j}{\partial \, w_i}
\,=\, \nabla_v \, \lambda_i \cdot {\bf e}_i \,=\, 0 \;\;.
\label{eq:degeneracy1} \end{equation}

Since in our case the \,$\lambda$'s\, don't depend on the \,$v$'s\,, it
is obvious that all the eigenfields are linearly degenerate.

In the case of systems of conservation laws where the sources vanish,
linear degeneracy is enough to guarantee that no shocks will form.
However, when the sources are non-zero, this is not true anymore.  This
is easy to see if we consider for a moment the prototype of non-linear
hyperbolic equations, Burgers' equation:  \begin{equation} \partial_t
\, u \,+\, u \, \partial_x \, u \,=\, 0 \;\;.  \label{eq:Burgues}
\end{equation}

If we now define:  \begin{equation} v \,:=\, \partial_x \, u \;\;,
\end{equation}

\noindent then we can rewrite equation~(\ref{eq:Burgues}) as the
system: \begin{mathletters} \begin{eqnarray} & \partial_t \, u \,=\,
-\, u \, v \;\;, & \\ & \partial_t \, v \,+\, \partial_x \, \left( \, u
\, v \, \right) \,=\, 0 \;\;. & \end{eqnarray} \end{mathletters}

\noindent This has precisely the form~(\ref{eq:system}).  The only
eigenvalue turns out to be equal to \,$u$\, which is clearly
independent of \,$v$.  By the definition above the system is linearly
degenerate.  However, it is clearly non-linear and will generate shocks
since it is only Burgues' equation in disguise.  The nonlinearities
have now been buried in the sources.

Clearly the condition that must be imposed to guarantee that no shocks
will develop is that a given eigenvalue \,$\lambda_i$\, should not be
affected by changes in the corresponding eigenfield \,$w_i$.  The
condition for linear degeneracy~(\ref{eq:degeneracy1}) asks for the
eigenvalue not to be explicitly dependent on its associated
eigenfield.  In the presence of sources, however, the coupling can
introduce an indirect dependency.  In order to study this dependency
let us  consider the time evolution of \,$\lambda_i$\,:
\begin{equation} \dot{\lambda_i} \,=\, \partial_t \, \lambda_i \,=\,
\sum_{j=1}^{N_u} \; \frac{\partial \, \lambda_i}{\partial \, u_j} \;
\partial_t \, u_j \,=\, \nabla_u \, \lambda_i \cdot {\bf p} \;\;.
\end{equation}

Now, we want this time derivative to be independent of the eigenfield
\,$w_i$:  \begin{equation} \frac{\partial \, \dot{\lambda_i}}{\partial
\, w_i} \,=\, \frac{\partial}{\partial \, w_i} \; \left( \nabla_u \,
\lambda_i \,\cdot\, {\bf p} \right) \,=\, 0 \;\;.
\label{eq:degeneracy2} \end{equation}

I shall call this condition `indirect linear degeneracy' and I will
refer to condition~(\ref{eq:degeneracy1}) as `explicit' or `direct'
linear degeneracy.

If we assume that the condition for explicit linear degeneracy holds,
then the condition for indirect linear degeneracy can be reduced to:
\begin{equation} \nabla_u \, \lambda_i \,\cdot\, \frac{\partial \, {\bf
p}}{\partial \, w_i} \,=\, \nabla_u \, \lambda_i \,\cdot\,
\sum_{j=1}^{N_v} \; \frac{\partial \, {\bf p}}{\partial \, v_j} \;
\frac{\partial \, v_j}{\partial \, w_i} \,=\, 0 \;\; , \end{equation}

\noindent which can be rewritten as:  \begin{equation} \nabla_u \,
\lambda_i \,\cdot\, \left( \rule{0mm}{4mm} {\bf e}_i \,\cdot\, \nabla_v
\, \right) \, {\bf p} \,=\, 0 \;\;. \end{equation}

This condition must supplement the condition for explicit linear
degeneracy~(\ref{eq:degeneracy1}) if we want to guarantee that no
shocks will develop.

\vspace{5mm}

Let us now apply the previous condition to the BM system of evolution
equations~(\ref{eq:evolution1}) and~(\ref{eq:evolution2}).  From the
discussion of the previous section it is clear that, on a given spatial
direction \,$x$\, we only have the following non-trivial eigenvalues:
\begin{equation} {\lambda^l}_\pm \,=\, \pm \, \alpha \,
\sqrt{\rule{0mm}{3mm} g^{xx}} \;\;, \qquad {\lambda^f}_\pm \,=\, \pm \,
\alpha \, \sqrt{f \, g^{xx}} \;\;.  \label{eq:lambdas} \end{equation}

The time derivative of \,${\lambda^l}_\pm$\, will then be:
\begin{eqnarray} \dot{\lambda^l}_\pm &=& \pm \, {\lambda^l}_\pm \;
\left[ \frac{1}{\alpha} \; \partial_t \, \alpha \,+\, \frac{1}{2 \,
g^{xx}} \; \partial_t \, g^{xx} \right] \nonumber \\ &=& \pm \,
{\lambda^l}_\pm \; \left[ \frac{1}{\alpha} \; \partial_t \, \alpha
\,-\, \frac{g^{xm} \, g^{xn}}{2 \, g^{xx}} \; \partial_t \, g_{mn}
\right] \;\; .  \end{eqnarray}

\noindent Using now equations~(\ref{eq:evolution1}) we find:
\begin{equation} \dot{\lambda^l}_\pm \,=\, \pm\, \alpha \,
{\lambda^l}_\pm \; \left( \rule{0mm}{4mm} K^{xx} / g^{xx} \,-\, f \,
{\rm tr} K \right) \;\; .  \end{equation}

\noindent Now, from the definitions of \,$w_l$\, and \,$w_f$\,
(equations~(\ref{eq:wl}) and~(\ref{eq:wf})) we can easily find that:
\begin{equation} {\rm tr} K \,=\, \frac{1}{2 \, \sqrt{f}} \, \left(
\rule{0mm}{4mm} {w^f}_+ \,+\, {w^f}_- \right) \;\; , \\ \end{equation}

\noindent and \,$( p,q \,\neq\, x)$:  \begin{eqnarray} K^{xx} &=&
g^{xx} \, {\rm tr} K \,+\, K_{pq} \, \left( \rule{0mm}{4mm} g^{xp} \,
g^{xq} \,-\, g^{xx} \, g^{pq} \right) \nonumber \\ &=& {\displaystyle
\frac{1}{2}} \; \left[ \frac{g^{xx}}{\sqrt{f}} \, \left(
\rule{0mm}{4mm} {w^f}_+ \,+\, {w^f}_- \right) \,+\, \left(
\rule{0mm}{4mm} g^{xp} \, g^{xq} \,-\, g^{xx} \, g^{pq} \right) \left(
w^l_{pq+} \,+\, w^l_{pq-} \right) \right] \;\; .  \end{eqnarray}

\noindent Substituting these results back in the expression for
\,$\dot{\lambda^l}_\pm$\, we find:  \begin{eqnarray}
\dot{\lambda^l}_\pm &=& \pm\, \frac{\alpha \, {\lambda^l}_\pm}{2} \;
\left[ \rule{0mm}{4mm} \frac{1}{\sqrt{f}} \, \left( \rule{0mm}{4mm} 1
\,-\, f \right) \left( \rule{0mm}{4mm} {w^f}_+ \,+\, {w^f}_- \right)
\right. \nonumber \\ && \left. \rule{0mm}{6mm} +\, \left(
\rule{0mm}{4mm} g^{xp} \, g^{xq} \,-\, g^{xx} \, g^{pq} \right) \left(
w^l_{pq+} \,+\, w^l_{pq-} \right) \right] \;\; .  \label{eq:lambdaldot}
\end{eqnarray}

In the same way we find for the time derivative of
\,${\lambda^f}_\pm$:  \begin{equation} \dot{\lambda^f}_\pm \,=\, \pm\,
\alpha \, {\lambda^f}_\pm \; \left[ \rule{0mm}{5mm} K^{xx} / g^{xx}
\,-\, \left( \rule{0mm}{4mm} f \,+\, \alpha \, f'/ 2 \right) \, {\rm
tr} K \right] \;\; , \end{equation}

\noindent where \,$f' \,=\, \partial_{\alpha} \, f$.  Substituting
again the expression for \,$K_{xx}$\, and \,${\rm tr} K$\, in terms of
the eigenfields we find:  \begin{eqnarray} \dot{\lambda^f}_\pm &=&
\pm\, \frac{\alpha \, {\lambda^f}_\pm}{2} \; \left[ \rule{0mm}{4mm}
\frac{1}{\sqrt{f}} \, \left( \rule{0mm}{4mm} 1 \,-\, f \,-\, \alpha \,
f' / 2 \right) \left( \rule{0mm}{4mm} {w^f}_+ \,+\, {w^f}_- \right)
\right. \nonumber \\ && \left. \rule{0mm}{6mm} +\, \left(
\rule{0mm}{4mm} g^{xp} \, g^{xq} \,-\, g^{xx} \, g^{pq} \right) \left(
w^l_{pq+} \,+\, w^l_{pq-} \right) \right] \;\; .  \label{eq:lambdafdot}
\end{eqnarray}

Equations~(\ref{eq:lambdaldot}) and~(\ref{eq:lambdafdot}) are very
important results.  Consider first the situation for
\,${\lambda^f}_\pm$.  If we want \,$\dot{\lambda^f}_\pm$ to be
independent of \,$w^f_\pm$\,, and hence satisfy the condition for
indirect linearly degeneracy, we must clearly ask for:
\begin{equation} 1 \,-\, f \,-\, \alpha \, f' / 2 \,=\ 0 \;\; .
\end{equation}

\noindent This differential equation can be easily solved to give:
\begin{equation} f \left( \alpha \right) \,=\, 1 \,+\, k \,/\, \alpha^2
\;\; , \label{eq:fcondition} \end{equation}

\noindent with \,$k$\, an arbitrary constant.  We must in fact take
\,$k \,\geq\, 0 $\, in order to ensure that we will have \,$f \,>\,
0$\, for all \,$\alpha \,>\, 0 $\,.

We have then show that the function \,$f$\, must have the
form~(\ref{eq:fcondition}) in order to guarantee that the eigenfields
\,$w^f_\pm$\, will not generate shocks.  Notice that if we take \,$k =
0$\, the condition reduces to that of harmonic slicing, {\em i.e.} for
harmonic slicing the eigenfields \,$w^f_\pm$\, do not generate shocks.

Consider now the situation for \,${\lambda^l}_\pm$.  From
equation~(\ref{eq:lambdaldot}) it is clear that if we want
\,$\dot{\lambda^l}_\pm$ to be independent of \,$w^l_{qp\pm}$\, we must
have:  \begin{equation} g^{xp} \, g^{xq} \,-\, g^{xx} \, g^{pq} \,=\, 0
\qquad \left( p,q \,\neq\, x \right) \;\;.  \end{equation}

This condition is very restrictive.  In particular, it is impossible to
satisfy with a diagonal metric.  We then reach the conclusion that in
the general case, the eigenfields \,$w^l_{qp\pm}$\, can {\em always}\/
generate shocks.  Notice how this result is independent of the value of
\,$f$, it will therefore remain true even in the case of harmonic
slicing.

One must stress here the fact that we haven't actually proved that shocks
will indeed develop.  Whether they do or not in any particular case should
depend in a critical way on the form of the initial data.

\vspace{5mm}

In the following sections I will consider some examples that show how
coordinate shocks can indeed develop even in very simple cases.

\section{FLAT TWO-DIMENSIONAL SPACETIME}

\subsection{Evolution equations}

As a first example, consider a flat two-dimensional spacetime (a `1+1'
spacetime) with coordinates \,$\{t,x\}$\,.  Notice that these coordinates
do not have to correspond to the Minkowski coordinates \,$\{x_M,t_M\}$\,,
so we can have a non-trivial evolution even though the spacetime is flat.
Since we only have one spatial dimension, I will simplify the notation
in the following way:
\begin{equation} g \,:=\, g_{xx} \; , \quad A
\,:=\, A_{x} \;, \quad D \,:=\, D_{xxx} \;, \quad K \,:=\, K_{xx}
\;. \end{equation}

\noindent Notice that the variable \,$V_x$\, is identically
zero.

The system of evolution equations~(\ref{eq:evolution1})
and~(\ref{eq:evolution2}) reduces in this case to:  \begin{mathletters}
\begin{eqnarray} \partial_t \, \alpha &=& -\, \alpha^2 \, f \, K / g
\;\;, \\ \partial_t \; g &=& -\, 2 \, \alpha \, K \;\;, \end{eqnarray}
\end{mathletters}

\noindent and: \begin{mathletters} \begin{eqnarray} \partial_t \, A
\,+\, \partial_x \left( \rule{0mm}{4mm} \alpha \, f \, K / g \right)
&=& 0 \;\;, \\ \partial_t \, D \,+\, \partial_x \left( \rule{0mm}{4mm}
\alpha \, K \right) &=& 0 \;\;, \\ \partial_t \, K \,+\, \partial_x \,
\left( \rule{0mm}{4mm} \alpha \, A \right) &=& \alpha / g \, \left( A
\, D \,-\, K^2 \right) \;\;. \end{eqnarray} \end{mathletters}

The characteristic structure of this system is very simple:

\begin{itemize}

\item There are 3 fields that propagate along the time lines (speed
zero).  These fields are:  \begin{equation} \left\{ \rule{0mm}{4mm}
\alpha,\, g,\, A  - f \, D / g \right\} \;\;.  \end{equation}

\item The 2 remaining fields propagate with the `gauge speeds':
\begin{equation} {\lambda^f}_\pm \,=\, \pm \, \alpha \, \sqrt{f / g}
\;\; .  \end{equation}

They are:  \begin{equation} {w^f}_\pm \,=\, \sqrt{f} \; K / g \,\pm\,
A / \sqrt{\rule{0mm}{3mm} g} \;\; . \end{equation}

\end{itemize}

Notice how there are no fields propagating along the physical light
cones.  According to the discussion of the previous section, we should
then expect shocks only when condition~(\ref{eq:fcondition}) is
violated.

\subsection{Numerical simulations}

Since we are dealing with a flat spacetime, the only way to obtain a
non-trivial evolution is to start with a non-trivial initial slice.  I will
therefore consider an initial slice given in terms of Minkowski
coordinates \,$\{x_M,t_M\}$\,as:  \begin{equation} t_M \,=\, h \left(
x_M \right) \;\; .  \end{equation}

I will assume that the dynamical spatial coordinate \,$x$\, coincides
initially with the Minkowski spatial coordinate \,$x_M$.  It is then
not difficult to show that the initial metric \,$g$\, and extrinsic
curvature \,$K$\, are given by: \begin{mathletters} \begin{eqnarray} g
& = & 1 \,-\, {h'\,}^2 \;\;, \\ K & = & -\, h'' / \sqrt{g} \;\;.
\end{eqnarray} \end{mathletters}

The initial value of \,$D$\, can be obtained directly from its
definition in terms of \,$g$.  The initial lapse is taken to be equal
to 1 everywhere, which implies that \mbox{\,$A \,=\, 0$}.

In all the simulations shown here, the function \,$h(x)$\, has a
Gaussian profile:  \begin{equation} h \left( x \right) \,=\, H \; \exp
\left\{ -\, \frac{\left( x - x_c \right)^2}{\sigma^2} \, \right\} \;\;,
\end{equation}

\noindent with \,$\{ H, \sigma, x_c \}$\, constants. The particular
values of \,$\{ H, \sigma \}$\, used in the simulations presented here
are:  \begin{equation} H \,=\, 5 \;\;, \qquad \sigma \,=\, 10 \;\;.
\end{equation}

\begin{figure}[h] \def\epsfsize#1#2{0.8#1}
\centerline{\epsfbox{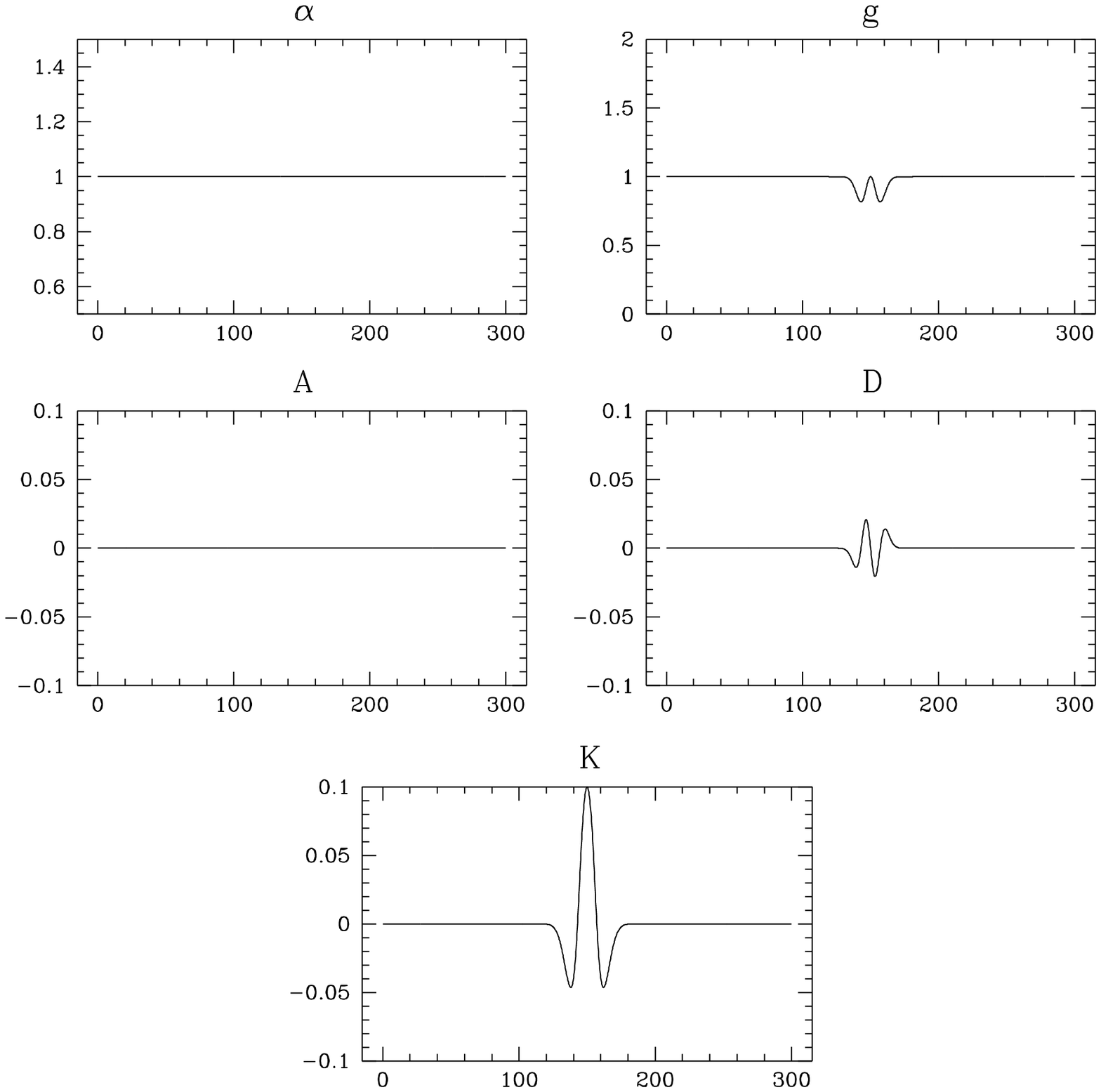}} \vspace{30mm} \caption{Two-dimensional
flat spacetime. Initial values of the dynamical variables.}
\label{fig:1+1_initial} \end{figure}

\noindent  I have also always taken the initial perturbation to be
centered around \mbox{\,$x_c \,=\, 150$}.  The initial values of all the
variables can be seen in Figure~\ref{fig:1+1_initial}.  All the results
presented below where obtained using a time step of \mbox{\,$\Delta t
\,=\, 0.125$\,} and a spatial increment of \mbox{\,$\Delta x \,=\,
0.25$}.

In all the simulations, the evolution proceeds at first in a similar
way:  The initial perturbation in \,$g$\,, \,$D$\, and \,$K$\, gives
rise to perturbations in \,$\alpha$\, and \,$A$\,.  These perturbations
rapidly develop into two separate pulses traveling in opposite
directions with a speed \,$\sim \sqrt{f}$\,.  What happens later
depends crucially on the form of the function \,$f(\alpha)$.

\begin{figure}[h] \def\epsfsize#1#2{0.8#1}
\centerline{\epsfbox{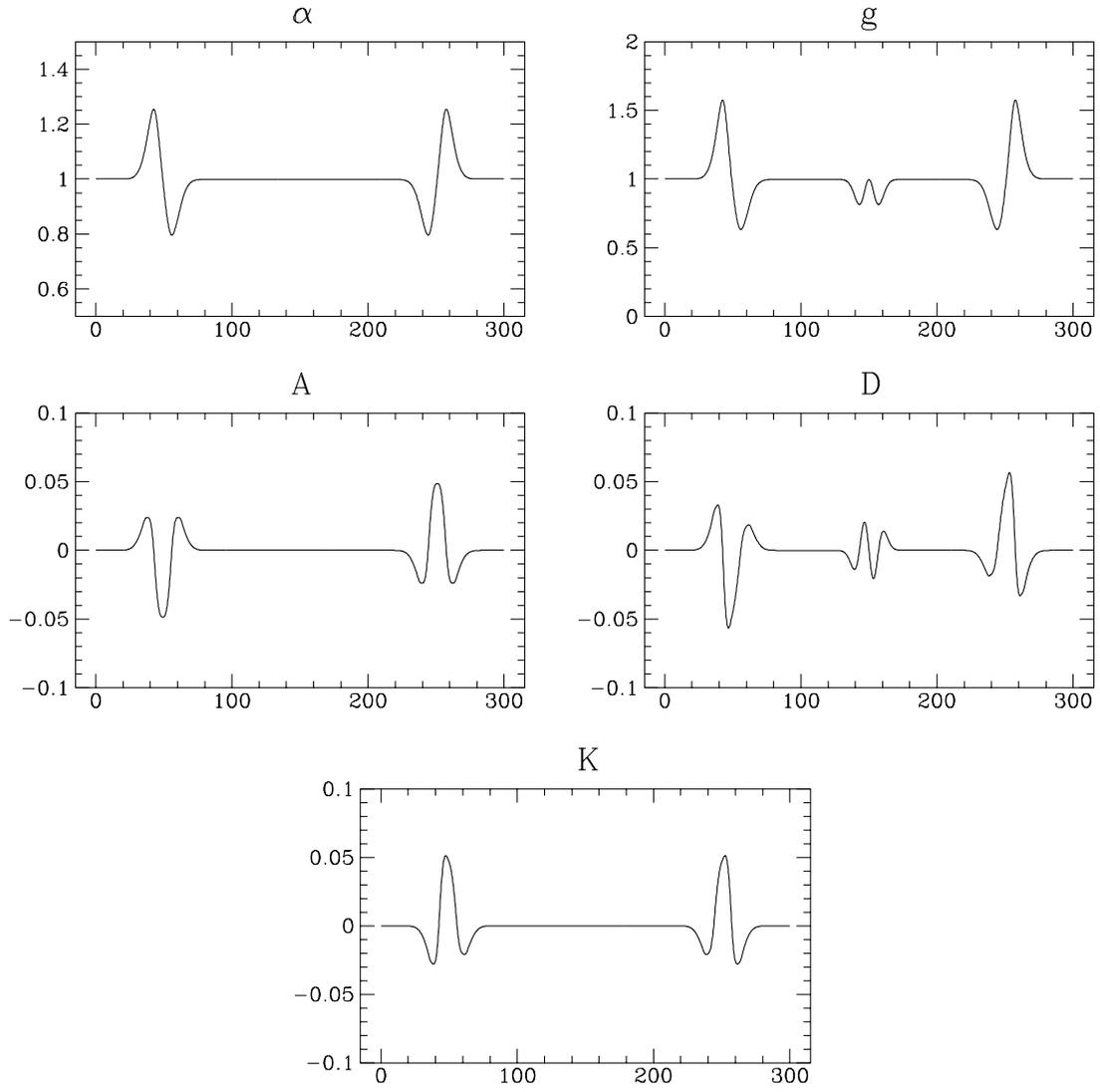}} \vspace{30mm} \caption{Two-dimensional
flat spacetime. Values of the variables at \,$t \,=\, 100$\, for harmonic
slicing \mbox{\,$(f \,=\, 1)$}.} \label{fig:1+1_f=1} \end{figure}

For harmonic slicing ($f \,=\, 1$), the pulses remain smooth as they
move away.  Once the pulses are gone, the lapse, the metric, and the
variables \,$A$\, and \,$D$\, return to their initial values, and the
extrinsic curvature becomes \,0.  Figure~\ref{fig:1+1_f=1} shows the
values of the variables at \mbox{\,$t \,=\, 100$\,}.

\begin{figure}[h] \def\epsfsize#1#2{0.8#1}
\centerline{\epsfbox{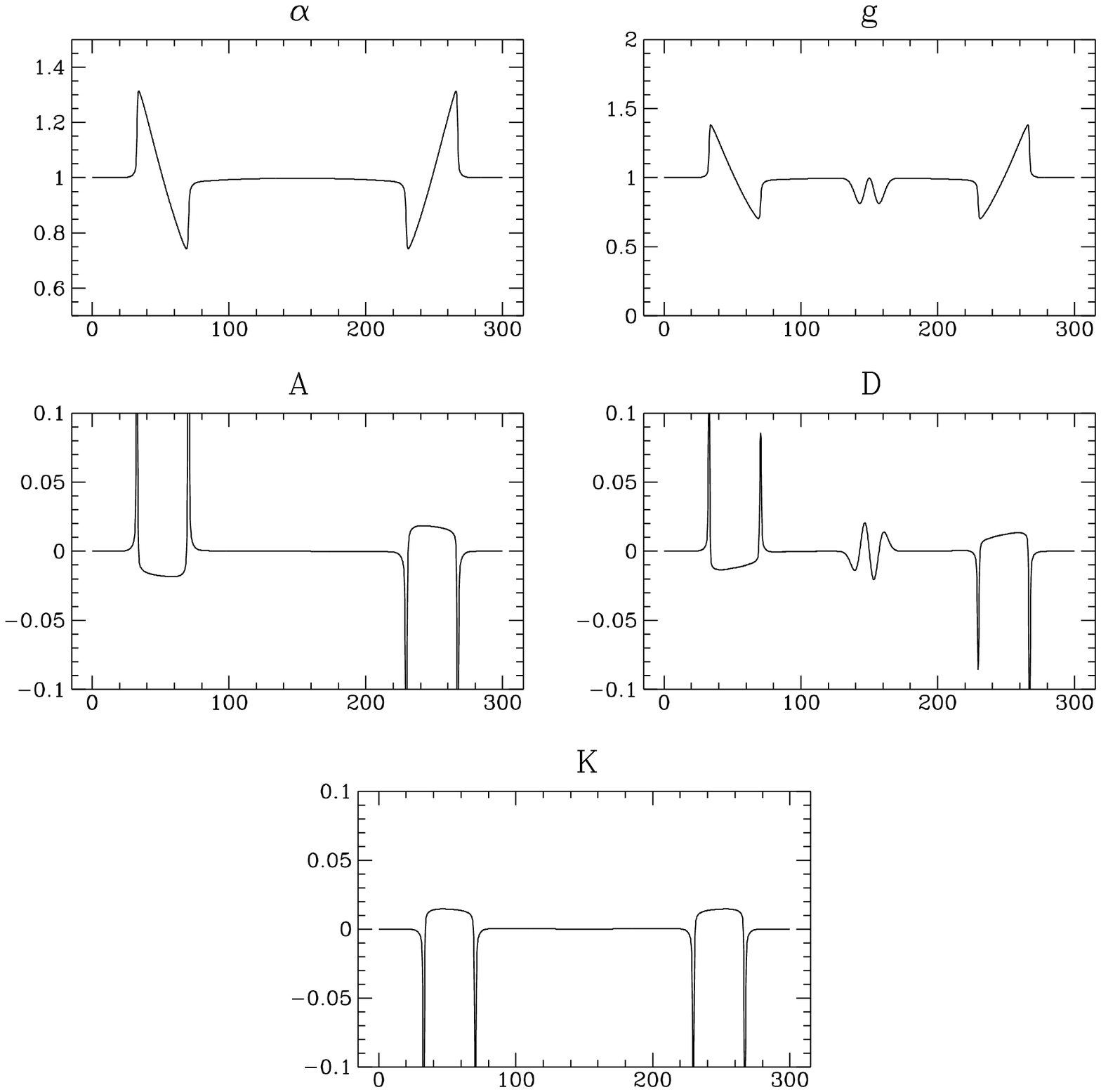}} \vspace{30mm} \caption{Two-dimensional
flat spacetime. Values of the variables at \,$t \,=\, 75$\, in the case when
\mbox{\,$f \,=\, 1.69$}.} \label{fig:1+1_f>1} \end{figure}

When \,$f$\, is a constant larger than 1, the pulses do not remain
smooth and shocks develop.  In fact, we have two shocks developing in
each pulse, one in front of it and one behind it.  At those points, the
lapse and the metric develop large gradients, while the extrinsic
curvature and the variables \,$A$\, and \,$D$\, develop very tall and
narrow spikes.  Figure~\ref{fig:1+1_f>1} shows the values of the
variables at \mbox{\,$t \,=\, 75$\,} in the particular case when
\mbox{\,$f \,=\, 1.69$}.

\begin{figure}[h] \def\epsfsize#1#2{0.8#1}
\centerline{\epsfbox{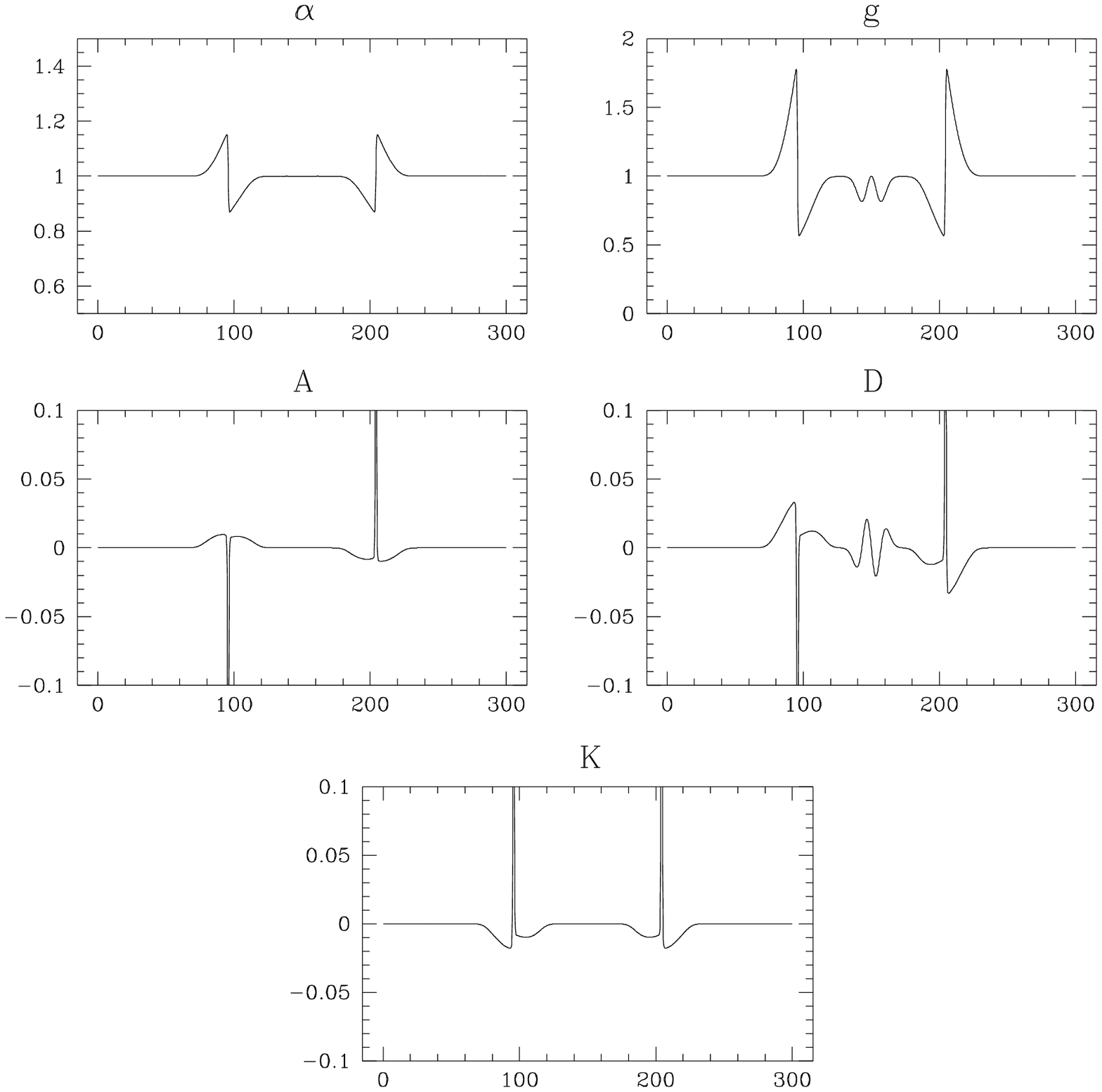}} \vspace{30mm} \caption{Two-dimensional
flat spacetime. Values of the variables at \,$t \,=\, 75$\, in the case
when \mbox{\,$f \,=\, 0.49$}.} \label{fig:1+1_f<1} \end{figure}

When \,$f$\, is a constant smaller than 1, a single shock develops in
the middle of each pulse.  Appart from this, the situation is very
similar to the case \mbox{\,$f \,>\, 1$}.  Figure~\ref{fig:1+1_f<1} shows
the values of the variables at \mbox{\,$t \,=\, 75$\,} in the particular
case when \mbox{\,$f \,=\, 0.49$}.

\begin{figure}[h] \def\epsfsize#1#2{0.8#1}
\centerline{\epsfbox{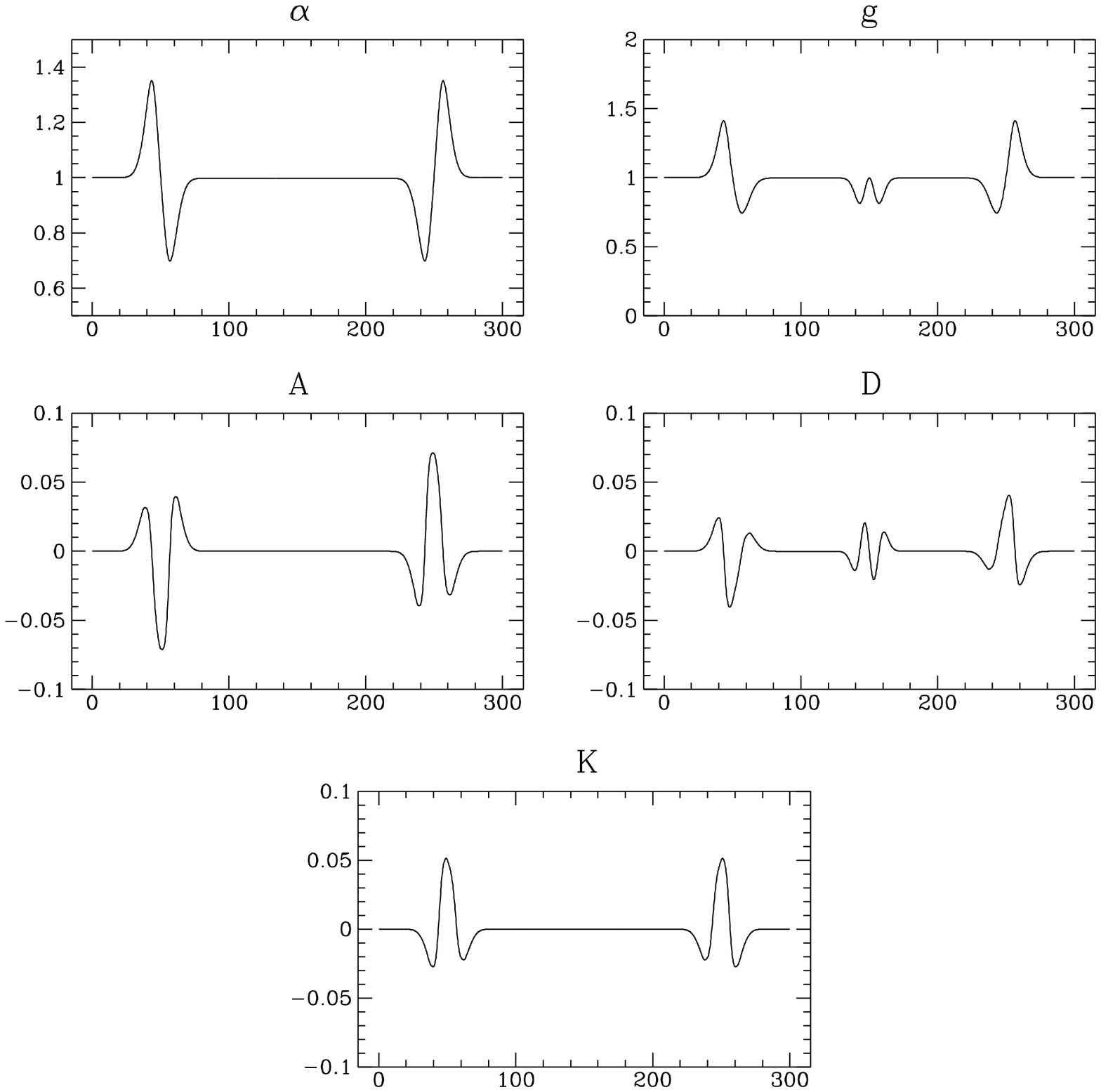}} \vspace{30mm} \caption{Two-dimensional
flat spacetime. Values of the variables at \,$t \,=\, 70$\, in the case when
\mbox{\,$f \,=\, 1 \,+\, 1 \,/\, \alpha^2$}.} \label{fig:1+1_f=1+1/a^2}
\end{figure}

Finally, when \,$f$\, is of the form~(\ref{eq:fcondition}), no shocks
develop in agreement with the predictions.  The pulses remain smooth and
move away with a speed \,$\sim \sqrt{1+k}$.
Figure~\ref{fig:1+1_f=1+1/a^2} shows the values of the variables at
\mbox{\,$t \,=\, 70$\,} in the particular case when
\,\mbox{$f \,=\, 1 \,+\, 1 \,/\, \alpha^2$}\,.

As a final comment, it should be mentioned that I have performed similar
simulations with many different values of the amplitude \,$H$\, and
width \,$\sigma$\, of the gaussian profile of the initial slice.
When \,$f$\, is not of the form~(\ref{eq:fcondition}) shocks apparently
always develop, though at different times.  The crucial feature that
seems to determine the time of shock formation is the maximum absolute
value of the extrinsic curvature \,$K$.  For small values of \,$K$\,,
shocks take a long time to appear, whereas for large values they develop
very rapidly.

\section{SPHERICALLY SYMMETRIC VACUUM SPACETIME}

\subsection{Evolution equations}

As a second example, consider a spherically symmetric four-dimensional
vacuum spacetime.  Let us introduce the coordinate system \,$\{t, r, \theta,
\phi \}$\,.  The only independent dynamical variables will then be:
\begin{equation} \left\{ \rule{0mm}{4mm} \alpha,\, g_{rr},\,
g_{\theta\theta},\, A_r,\, D_{rrr},\, D_{r \theta \theta},\, K_{rr},\,
K_{\theta \theta},\, V_r \right\} \;\; .  \end{equation}

The system of evolution equations~(\ref{eq:evolution1})
and~(\ref{eq:evolution2}) reduces now to: \begin{mathletters}
\begin{eqnarray} \partial_t \, \alpha &=& -\, \alpha^2 \, f \, {\rm tr}
K \;\;, \\ \partial_t \; g_{rr} &=& -\, 2 \, \alpha \, K_{rr} \;\;,
\\ \partial_t \; g_{\theta\theta} &=& -\, 2 \, \alpha \,
K_{\theta\theta} \;\;, \end{eqnarray} \end{mathletters}

\noindent and:  \begin{mathletters} \begin{eqnarray} \partial_t \, A_r
\,+\, \partial_r \left( \rule{0mm}{4mm} \alpha \, f \, {\rm tr} K
\right) &=& 0 \;\;, \\ \rule{0mm}{6mm} \partial_t \, D_{rrr} \,+\,
\partial_r \left( \rule{0mm}{4mm} \alpha \, K_{rr} \right) &=& 0 \;\;,
\\ \rule{0mm}{6mm} \partial_t \, D_{r\theta\theta} \,+\, \partial_r
\left( \rule{0mm}{4mm} \alpha \, K_{\theta\theta} \right) &=& 0 \;\;,
\\ \rule{0mm}{6mm} \partial_t \, K_{rr} \,+\, \partial_r \, \left(
\rule{0mm}{4mm} \alpha \, \lambda^r_{rr} \right) &=& \alpha \, S_{rr}
\;\;, \\ \rule{0mm}{6mm} \partial_t \, K_{\theta\theta} \,+\,
\partial_r \, \left( \rule{0mm}{4mm} \alpha \, \lambda^r_{\theta\theta}
\right) &=& \alpha \, S_{\theta\theta} \;\;, \\ \rule{0mm}{6mm}
\partial_t \, V_r &=& \alpha \, P_r \;\;. \end{eqnarray}
\end{mathletters}

\noindent with:  \begin{mathletters} \begin{eqnarray} \lambda^r_{rr}
&=& A_r \,+\, 2 \, V_r \,-\, 2 \, D_{r\theta\theta} / g_{\theta\theta}
\;\;, \\ \lambda^r_{\theta\theta} &=& D_{r\theta\theta} / g_{rr} \;\;,
\end{eqnarray} \end{mathletters}

\noindent and:  \begin{mathletters} \begin{eqnarray} S_{rr} &=& K_{rr}
\, \left( \rule{0mm}{4mm} 2 \, K_{\theta\theta} / g_{\theta\theta}
\,-\, K_{rr} / g_{rr} \right)  \,+\, A_r \, \left( \rule{0mm}{4mm}
D_{rrr} / g_{rr} \,-\, 2 \, D_{r\theta\theta} / g_{\theta\theta}
\right) \\ && +\, 2 \, D_{r\theta\theta} / g_{\theta\theta} \left(
\rule{0mm}{4mm} D_{rrr} / g_{rr} \,-\, D_{r\theta\theta} /
g_{\theta\theta} \right) \,+\, 2 \, A_r \, V_r \;\;, \\ \rule{0mm}{7mm}
S_{\theta\theta} &=& K_{rr} \, K_{\theta\theta} / g_{rr} \,-\, D_{rrr}
\, D_{r\theta\theta} / {g_{rr}}^2 \,+\, 1 \;\;, \\ \rule{0mm}{7mm} P_r
&=& -\, 2 / g_{\theta\theta} \, \left[ \rule{0mm}{5mm} A_r \,
K_{\theta\theta} \,-\, D_{r\theta\theta} \, \left( \rule{0mm}{4mm}
K_{\theta\theta} / g_{\theta\theta} \,-\, K_{rr} / g_{rr} \right)
\right] \;\;. \end{eqnarray} \end{mathletters}

We also have the following algebraic constraint that must be satisfied
by the physical solutions:  \begin{equation} V_r \,=\, 2 \,
D_{r\theta\theta} / g_{\theta\theta} \;\; .  \end{equation}

The characteristic structure of this system turns out to be:

\begin{itemize}

\item 5 fields propagate along the time lines (speed zero).  These
fields are:  \begin{equation} \left\{ \rule{0mm}{4mm} \alpha,\,
g_{rr},\, g_{\theta\theta},\, V_r,\, A_r  - f \, {{D_r}^m}_m \right\}
\;\; .  \end{equation}

\item 2 fields propagate along the physical light cones with speeds:
\begin{equation} {\lambda^l}_\pm \,=\, \pm \, \alpha \,/ \sqrt{g_{rr}}
\;\; .  \end{equation}

These fields are:  \begin{equation} {w^l}_\pm \,=\, \sqrt{g_{rr}} \;
K_{\theta\theta} \,\pm\, D_{r\theta\theta} \;\; . \end{equation}

\item 2 fields propagate with the `gauge speeds':  \begin{equation}
{\lambda^f}_\pm \,=\, \pm \, \alpha \, \sqrt{f / g_{rr}} \;\; .
\end{equation}

They are:  \begin{equation} {w^f}_\pm \,=\, \sqrt{f \, g_{rr}} \; {\rm
tr} K \,\pm\, \left( \rule{0mm}{4mm} A_r \,+\, 2 \, V_r \right) \;\; .
\end{equation}

\end{itemize}

Notice how we now have both fields propagating with the speed of light
and fields propagating with the gauge speed.  We should then expect to
see two different types of shocks forming.  In particular, shocks
produced by the \,${w^l}_\pm$\, fields can be expected always, even
for harmonic slicing.

\subsection{Numerical simulations for a flat spacetime}

Again, since we are dealing with flat spacetime, the only way to obtain
a non-trivial evolution is to start with a non-trivial initial slice. I
will therefore consider an initial slice given in terms of Minkowski
coordinates \,$\{r_M,t_M\}$\,as:  \begin{equation} t_M \,=\, h \left(
r_M \right) \;\; .  \end{equation}

I will assume that the dynamical radial coordinate \,$r$\, coincides
initially with the Minkowski radial coordinate \,$r_M$.  It is then not
difficult to show that the initial metric
\,$\{g_{rr},g_{\theta\theta}\}$\, and extrinsic curvature
\,$\{K_{rr},K_{\theta\theta}\}$\, are given by:  \begin{mathletters}
\begin{eqnarray} g_{rr} & = & 1 \,-\, {h' \,}^2 \;\;,
\\ g_{\theta\theta} & = & r^2 \;\;, \\ K_{rr} & = & -\; h'' /
\sqrt{g_{rr}} \;\;, \\ K_{\theta\theta} & = & -\; r \, h' /
\sqrt{g_{rr}} \;\;. \end{eqnarray} \end{mathletters}

The initial values of \,$\{D_{rrr},D_{r\theta\theta},V_r\}$\, can be
obtained directly from their definitions in terms of the metric.  The
initial lapse is taken to be equal to 1 everywhere, which implies that
\mbox{\,$A_r \,=\, 0$}.

\begin{figure}[h] \def\epsfsize#1#2{0.9#1}
\centerline{\epsfbox{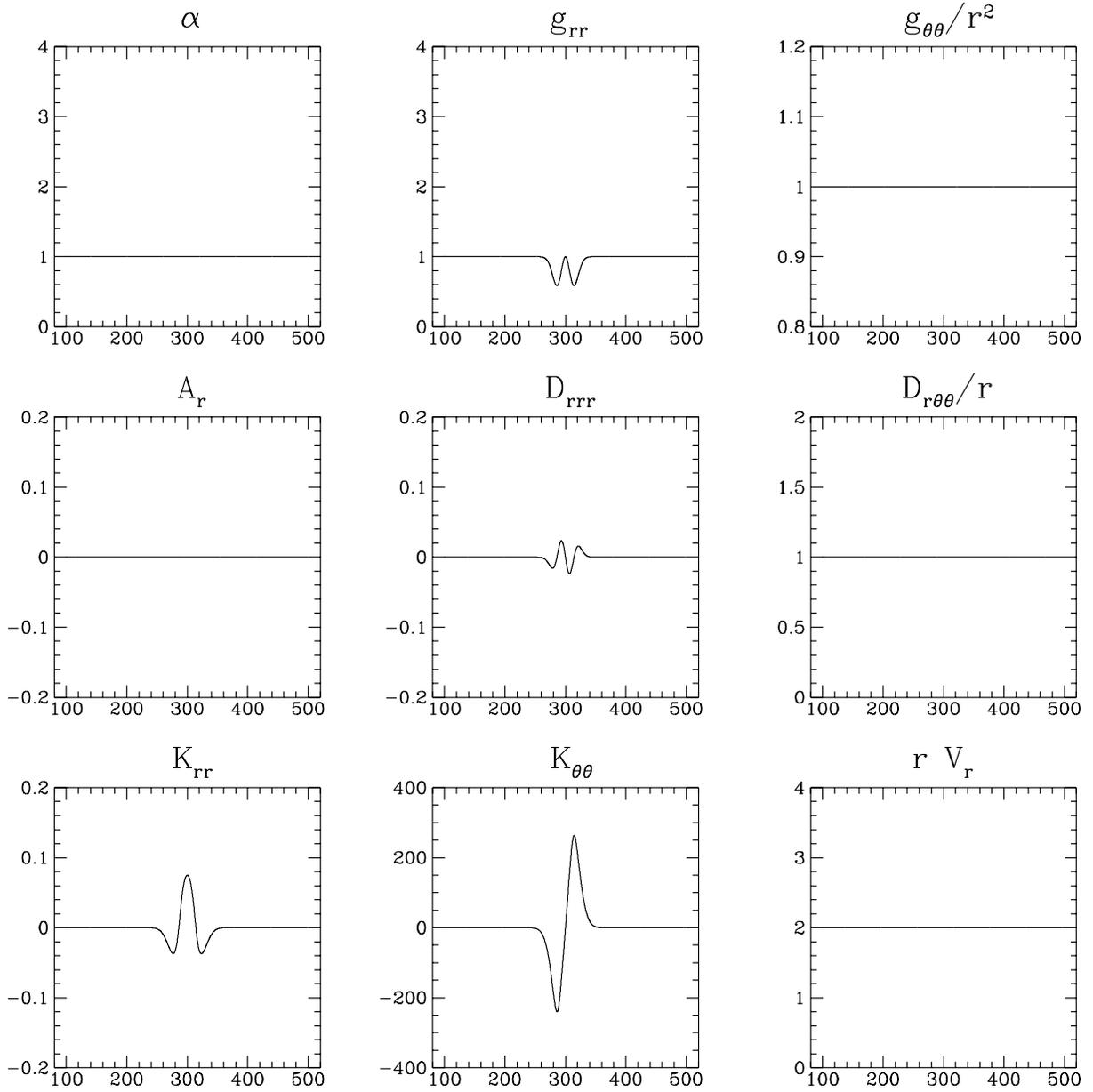}} \vspace{30mm} \caption{Spherically
symmetric flat spacetime. Initial values of the dynamical variables.}
\label{fig:SSF_initial} \end{figure}

In all the simulations shown here, the function \,$h(r)$\, has a
Gaussian profile:  \begin{equation} h \left( r \right) \,=\, H \; \exp
\left\{ -\, \frac{\left( r - r_c \right)^2}{\sigma^2} \, \right\} \;\;
, \end{equation}

\noindent with \,$\{ H, \sigma, r_c \}$\, constants.  The particular
values of \,$\{ H, \sigma \}$\, used in the simulations presented here
are:  \begin{equation} H \,= \, 15 \;\;, \qquad \sigma \,= \, 20 \;\;,
\end{equation}

\noindent  and I have taken the initial perturbation to be centered
around \mbox{\,$r_c \,=\, 300$}.  The initial values of all the
variables can be seen in Figure~\ref{fig:SSF_initial}.  The results
presented below where obtained using a time step of \mbox{\,$\Delta t
\,=\, 0.1$\,} and a spatial increment of \mbox{\,$\Delta x \,=\,
0.2$}.

In all the simulations the evolution proceeds at first in a similar way:
the initial perturbations in \,$\{g_{rr},\, D_{rrr},\, K_{rr},\,
K_{\theta\theta}\}$\, give rise to perturbations in \,$\{\alpha,\,
g_{\theta\theta},\, A_r,\, D_{r\theta\theta},\, V_r\}$.  These
perturbations develop into separate pulses traveling in opposite
directions with a speed \,$\sim \sqrt{f}$\,.  The pulses are not
symmetric any more since clearly the in going and out going directions
are not equivalent.

\begin{figure}[h] \def\epsfsize#1#2{0.9#1}
\centerline{\epsfbox{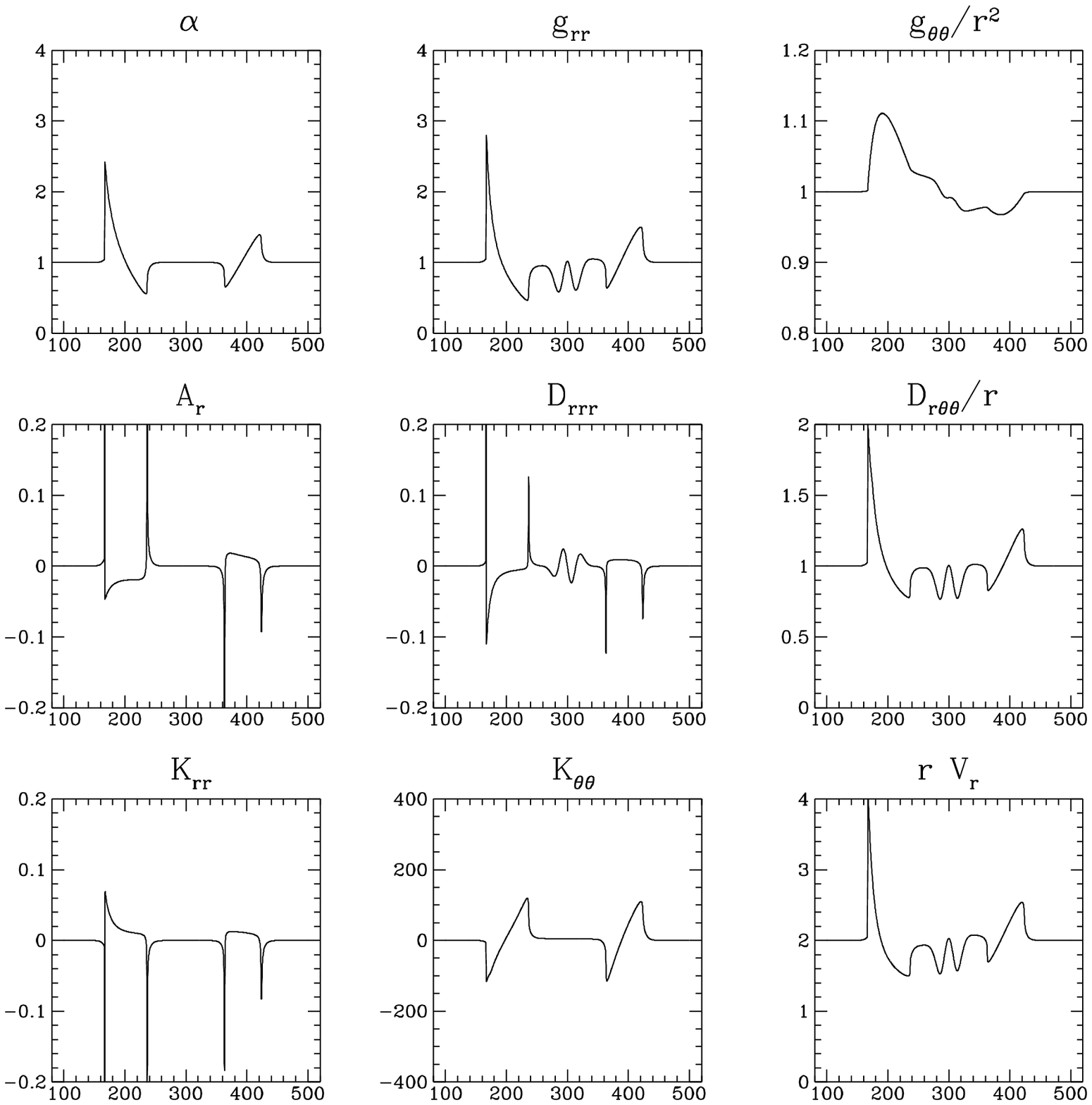}} \vspace{30mm} \caption{Spherically
symmetric flat spacetime. Values of the variables at \,$t \,=\, 70$\,
in the case when \mbox{\,$f \,=\, 1.69$}.} \label{fig:SSF_f>1}
\end{figure}

Consider first the case of \,$f > 1$.  As the evolution proceeds,
shocks develop in both pulses.  These shocks are similar to those found
in the 1+1 case:  two shocks develop in each
pulse, one in front of it and one behind it.  At those points
\,$\{\alpha,\, g_{rr},\, D_{r\theta\theta},\, K_{\theta\theta},\, V_r\}$\,
develop large gradients, while \,$\{A_r,\, D_{rrr},\, K_{rr}\}$\, develop
tall and narrow spikes.  The angular metric component \,$g_{\theta\theta}$\,
in contrast develops sharp corners.  Figure~\ref{fig:SSF_f>1} shows  the
values of the variables at \mbox{\,$t \,=\, 70$\,} in the particular
case when \mbox{\,$f \,=\, 1.69$}.

\begin{figure}[h] \def\epsfsize#1#2{0.9#1}
\centerline{\epsfbox{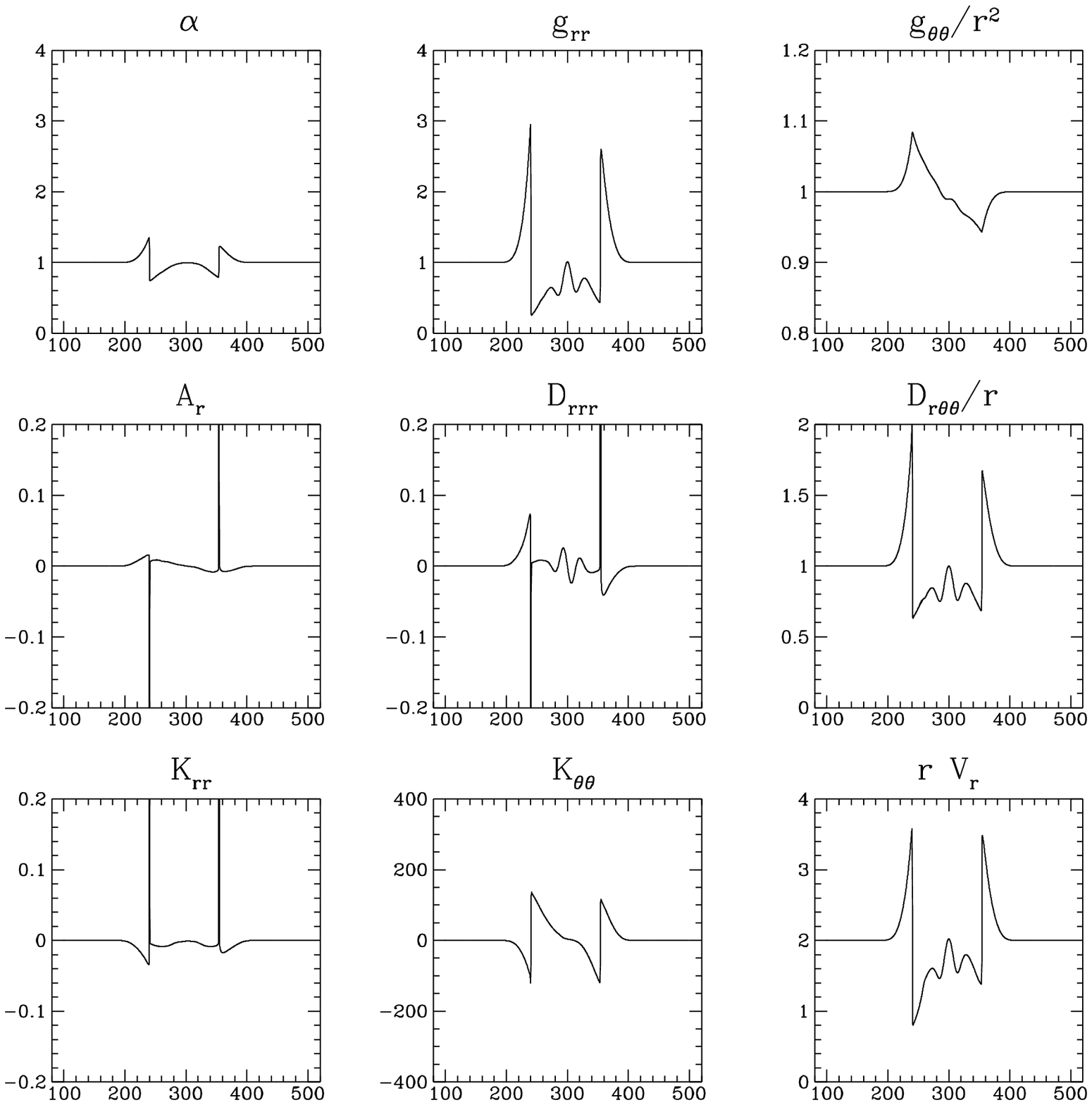}} \vspace{30mm} \caption{Spherically
symmetric flat spacetime. Values of the variables at \,$t \,=\, 70$\,
in the case when \mbox{\,$f \,=\, 0.49$}.} \label{fig:SSF_f<1}
\end{figure}

When \,$f < 1$, we again find results that are similar to the 1+1 case:
a single shock develops in each pulse.  Again, at the shock
\,$\{\alpha,\, g_{rr},\, D_{r\theta\theta},\, K_{\theta\theta},\,
V_r\}$\,  develop large gradients, \,$\{A_r,\, D_{rrr},\, K_{rr}\}$\,
develop spikes and \,$g_{\theta\theta}$\, develops sharp corners.
Figure~\ref{fig:SSF_f<1} shows  the values of the variables at
\mbox{\,$t \,=\, 70$\,} in the particular case when \mbox{\,$f \,=\,
0.49$}.

\begin{figure}[h] \def\epsfsize#1#2{0.9#1}
\centerline{\epsfbox{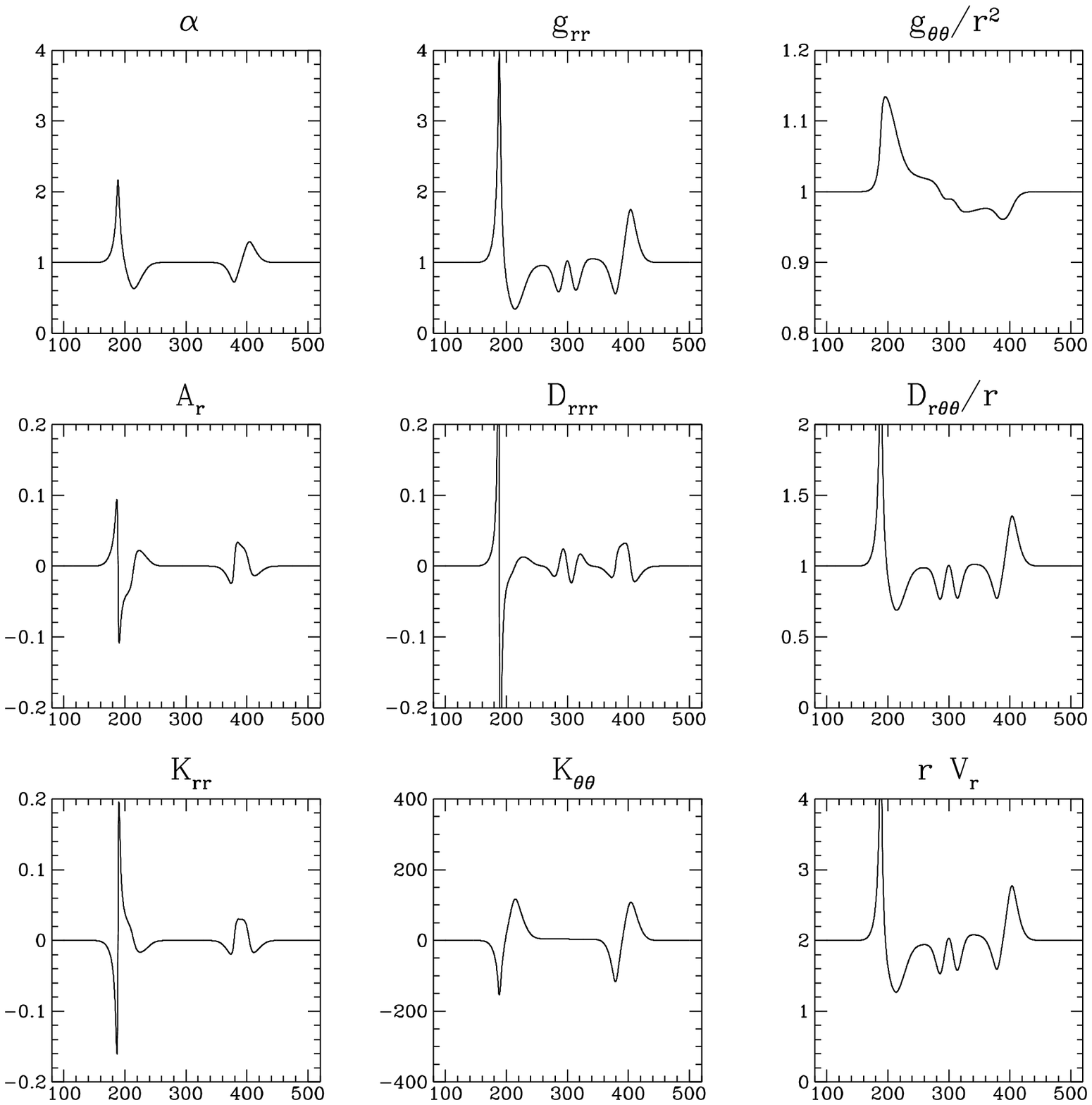}} \vspace{30mm} \caption{Spherically
symmetric flat spacetime. Values of the variables at \,$t \,=\, 70$\,
in the case when \mbox{\,$f \,=\, 1$}.} \label{fig:SSF_f=1} \end{figure}

The most interesting case is that of harmonic slicing ($f \,=\, 1$).  In
contrast to the 1+1 case, shocks still develop here.  The shocks,
however, have a different structure indicative of their different
origin: the variables \,$\{A_r,\, D_{rrr},\, K_{rr}\}$\, now develop
large gradients, while \,$\{\alpha,\, g_{rr},\, D_{r\theta\theta},\,
K_{\theta\theta},\, V_r\}$\, develop sharp spikes.  The angular metric
component \,$g_{\theta\theta}$\, also seems to develop a large
gradient, thought this gradient is less sharp than that found in other
variables.  This is easy to understand geometrically:  any
discontinuity in \,$g_{\theta\theta}$\, must necessarily be accompanied
by an infinite value of \,$g_{rr}$\, (we must jump a finite radial
distance in an infinitesimal interval). The shocks are clearly visible
in the in going pulse, but don't seem to be present in the out going pulse.
Figure~\ref{fig:SSF_f=1} shows the values of the variables at \mbox{\,$t
\,=\, 70$\,} for harmonic slicing.

Again, I have performed similar simulations for different amplitudes and
widths of the gaussian profile of the initial slice.  The shocks are
always there.  The only exception seems to be that for very small amplitudes
of the initial perturbation the pulses moving in the in going direction may
not have enough time to develop before they reach the origin.  Also,
for the case of harmonic slicing, shocks in the out going direction
never seem to form, no matter how large the initial perturbation might be.

\subsection{Numerical simulations for a black hole spacetime}

In all the previous examples I have restricted myself to a flat
spacetime.  Since this is a very special case one might think that the
shocks that we have found are just an artifact of the flatness.  To
show that this is not the case, I will now consider a spherically
symmetric black hole spacetime.

To find adequate initial data I start from a Schwarzschild slice with
spatial metric:  \begin{equation} d l^2 \,=\, \frac{1}{1 - 2 M/r_s} \;
d r_s^2 \,+\, r_s^2 \, d \Omega^2 \;\; , \end{equation}

\noindent where \,$r_s$\, is the Schwarzschild radial coordinate and
\mbox{\,$d \Omega^2 \,=\, d \theta^2 \,+\, {\sin}^2 \theta \, d
\phi^2$}.

In order to eliminate the singularity at \,$r_s = 2 M$, I will define a
new radial coordinate \,$r$\, that measures proper distance along the
slice.  The coordinates \,$r_s$\, and \,$r$\, will be related by:
\begin{equation} r \,=\, \eta \left( r_s \right) \,+\, M \, \ln \;
\left[ \frac{r_s + \eta(r)}{r_s - \eta(r_s)} \right] \;\;,
\label{eq:r(rs)} \end{equation}

\noindent with:  \begin{equation} \eta \left( r_s \right) \,=\, \left(
r_s^2 \,-\, 2 M r_s \right)^{1/2} \;\;.  \end{equation}

Notice that even though (\ref{eq:r(rs)}) can not be inverted
analytically to find \,$r_s(r)$, it can easily be inverted numerically
to arbitrarily high accuracy.

The new metric will now have the form:  \begin{equation} d l^2 \,=\, d
r^2 \,+\, \left( \rule{0mm}{4mm} r_s \left( r \right) \right)^2 \, d
\Omega^2 \;\;.  \end{equation}

It is easy to see that the Schwarzschild slice has zero extrinsic
curvature, so our initial data will be:  \begin{mathletters}
\begin{eqnarray} g_{rr} &=& 1 \;\;, \\ g_{\theta\theta} &=& r_s^2 \;\;,
\\ K_{rr} &=& 0 \;\;, \\ K_{\theta\theta} &=& 0 \;\;. \end{eqnarray}
\end{mathletters}

Now, if we use this initial data directly we will not see any shocks
develop.  This is known since the BM formalism has been used before to
solve this problem and no shocks have been
observed~\cite{BonaMasso95}.  The reason why shocks don't develop is
that they are a consequence of transport and as such they should only
develop when we have wave propagation, either in the form of real
gravitational waves, or in the form of pure gauge waves.  The static
black hole problem has no gravitational waves, and the initial data
given above does not give rise to gauge waves either.

In order to introduce gauge waves into our problem, I will consider an
initial slice given in terms of Schwarzschild time \,$t_s$\, in the
following way:  \begin{equation} t_s \,=\, h \left( r \right) \;\;.
\end{equation}

It is not difficult to show that the new slice will have the following
metric components:  \begin{mathletters} \begin{eqnarray} g_{rr} &=& 1
\,-\, \left( \rule{0mm}{4mm} \alpha_s \, h' \right)^2 \;\;,
\\ g_{\theta\theta} &=& r_s^2 \;\;, \end{eqnarray} \end{mathletters}

\noindent where \,$\alpha_s$\, is the Schwarzschild lapse function:
\begin{equation} \alpha_s \,=\, \left( \rule{0mm}{4mm} 1 \,-\, 2 M /
r_s \right)^{1/2} \;\;.  \end{equation}

The components of the extrinsic curvature for this slice can now be
shown to be:  \begin{mathletters} \begin{eqnarray} K_{rr} &=& -\;
\left[ \rule{0mm}{5mm} \alpha_s \, h'' \,+\, \alpha_s' \, h' \, \left(
2 \,-\, \left( \alpha_s \, h' \right)^2 \right) \right] / \sqrt{g_{rr}}
\;\;, \\ K_{\theta\theta} &=& -\; \alpha_s^2 \, r_s \, h' /
\sqrt{g_{rr}} \;\;. \end{eqnarray} \end{mathletters}

As before, the initial values of
\,$\{D_{rrr},D_{r\theta\theta},V_r\}$\, can be obtained directly from
their definitions in terms of the metric.  The initial lapse is again
taken to be equal to 1 everywhere, which implies that \mbox{\,$A_r
\,=\, 0$}.

For the function \,$h(r)$\, I will again use a Gaussian:
\begin{equation} h \left( r \right) \,=\, H \; \exp \left\{ -\,
\frac{\left( r - r_c \right)^2}{\sigma^2} \, \right\} \;\; ,
\end{equation}

\noindent with \,$\{ H, \sigma, r_c \}$\, constants.

In order to see the development of the shocks clearly, I will consider
simulations where the center of our perturbation \,$r_c$\, is out in
the wave zone.

All the simulations I have carried out proceed in a similar way.  At
the throat of the wormhole we find what we expect for a black hole
spacetime: the lapse collapses and the metric component \,$g_{rr}$\,
grows rapidly.  Out in the wave zone, the disturbance behaves in the
same way as it did in flat spacetime:  the initial perturbations in
\,$\{g_{rr},\, D_{rrr},\, K_{rr},\, K_{\theta\theta}\}$\, give rise to
perturbations in \,$\{\alpha,\, g_{\theta\theta},\, A_r,\,
D_{r\theta\theta},\, V_r\}$,  these then develop into separate pulses
traveling in opposite directions with a speed \,$\sim \sqrt{f}$\,.

In all cases, the traveling pulses develop shocks that have very
similar characteristics to those that we found in the flat case.  Here
I will only show the results found in the case of harmonic slicing \,$f
= 1$.  The particular values of \,$\{ H, \sigma \}$\, used in this
simulation are:  \begin{equation} H \,= \, 5 \;\;, \qquad \sigma \,=\,
5 \;\;.  \end{equation}

\begin{figure}[h] \def\epsfsize#1#2{0.9#1}
\centerline{\epsfbox{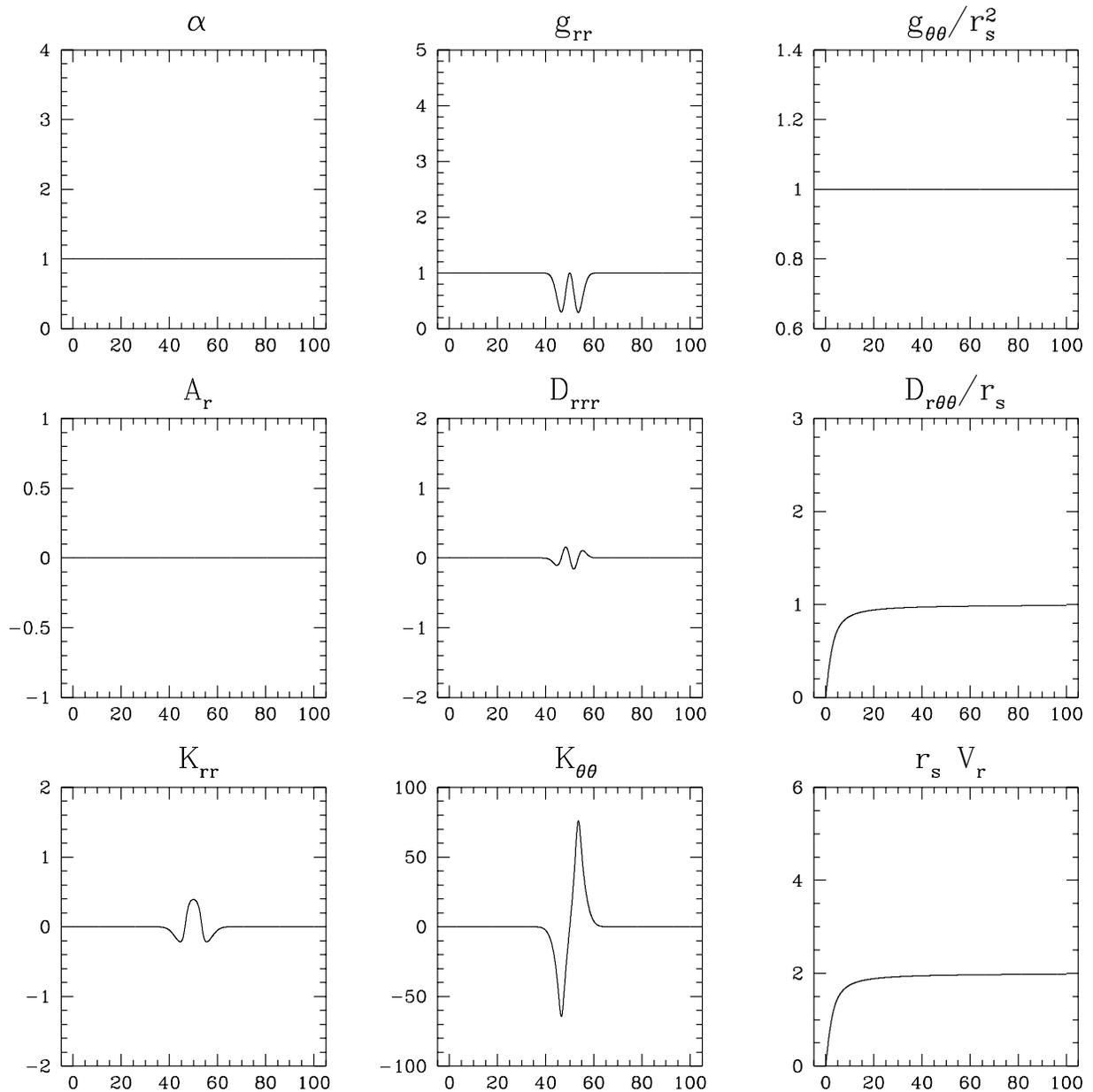}} \vspace{30mm} \caption{Spherically
symmetric black hole spacetime. Initial values of the dynamical
variables.} \label{fig:BH_initial} \end{figure}

\noindent  I have also taken the initial perturbation to be centered
around \mbox{\,$r_c \,=\, 50$}, and the mass of the black hole to be
\mbox{\,$M \,=\, 1$}.  The results presented here where obtained using
a time step of \mbox{\,$\Delta t \,=\, 0.025$\,} and a spatial
increment of \mbox{\,$\Delta x \,=\, 0.05$}. The initial values of all
the variables can be seen in Figure~\ref{fig:BH_initial}.

\begin{figure}[h] \def\epsfsize#1#2{0.9#1}
\centerline{\epsfbox{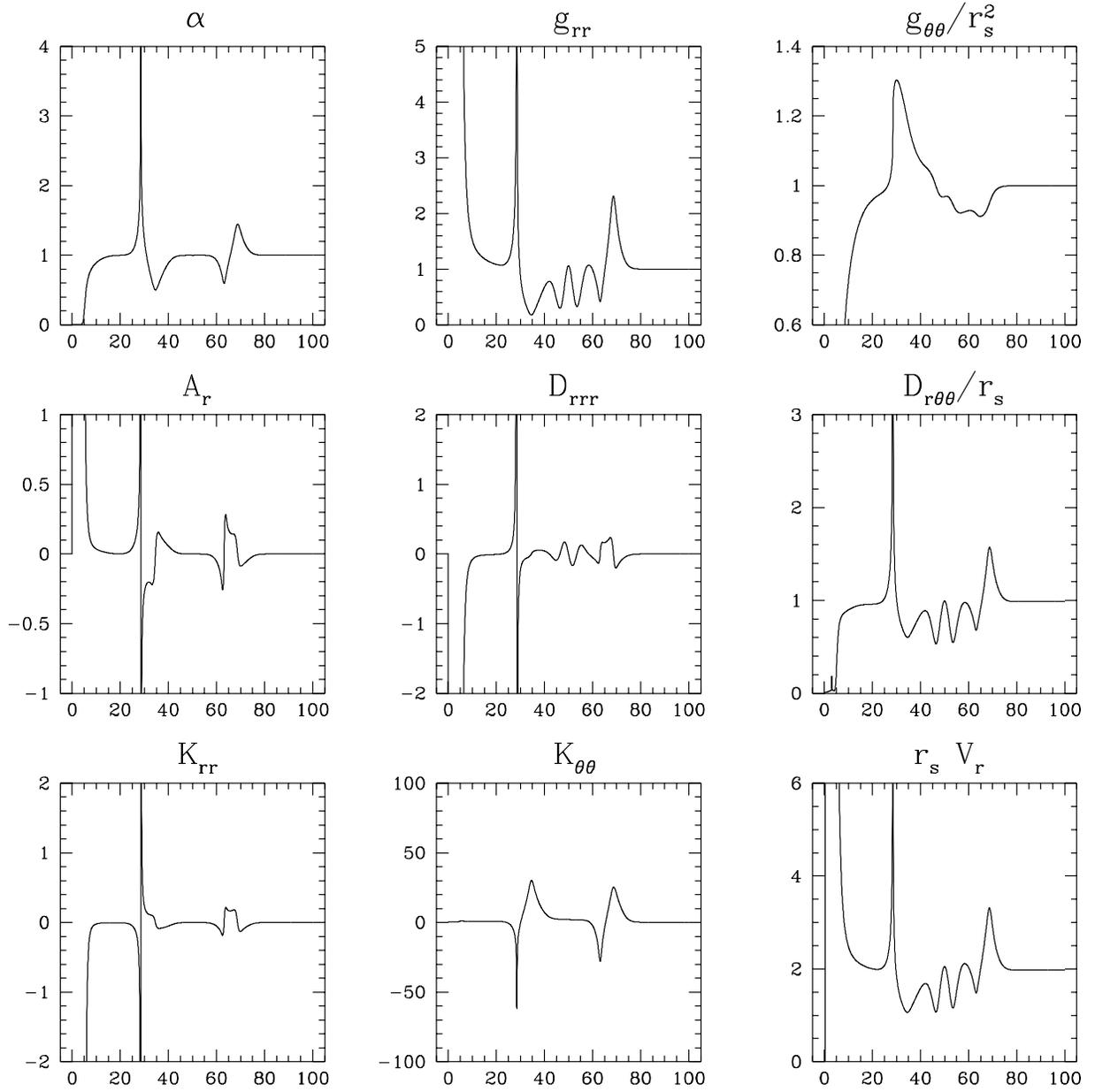}} \vspace{30mm} \caption{Spherically
symmetric black hole spacetime. Values of the variables at \,$t \,=\,
15$\, in the case when \mbox{\,$f \,=\, 1$}.} \label{fig:BH_f=1}
\end{figure}

Figure~\ref{fig:BH_f=1} shows the values of the variables at \mbox{\,$t
\,=\, 15$}.  Notice how around the throat the lapse and the angular
metric component \,$g_{\theta\theta}$\, have collapsed, while the
radial metric component \,$g_{rr}$\, has grown to a very large value.
The interesting region for our purposes, however, is away from the
throat.  We can clearly see the two pulses resulting from our initial
perturbation.  The pulse moving inwards has developed a shock:  the
variables \,$\{A_r,\, D_{rrr},\, K_{rr}\}$\, have developed large
gradients, while \,$\{\alpha,\, g_{rr},\, D_{r\theta\theta},\,
K_{\theta\theta},\, V_r\}$\, have developed sharp spikes. The angular
metric component \,$g_{\theta\theta}$\, has also developed a large
gradient.

\section{DISCUSSION}

I have introduced a general approach to the study of shock development
in hyperbolic systems of equations with sources.  I have shown that the
usual condition of explicit linear degeneracy (direct linear
degeneracy) must be supplemented with a new condition which I have
called `indirect linear degeneracy' in order to guarantee that no
shocks will develop.

I have applied this condition of indirect linear degeneracy to the BM
hyperbolic formalism of General Relativity in the case of a zero shift
vector.  My analysis has shown how two distinct families of
characteristic fields can give rise to shocks.  Numerical simulations
have confirmed these predictions in the simple cases of a flat
two-dimensional spacetime, a flat four-dimensional spacetime with
spherically symmetric slices, and a spherically symmetric black hole
spacetime.

The appearance of shocks that develop from smooth initial data in
vacuum General Relativity comes as a great surprise.  These shocks,
however, do not represent discontinuities in the geometry of spacetime,
but indicate instead regions where our coordinate system becomes
pathological.  It is for this reason that I refer to them as
`coordinate shocks'.

Of the two families of coordinate shocks found, one can be completely
eliminated by choosing a BM gauge function \,$f(\alpha)$\, of the
form:  \begin{equation} f \left( \alpha \right) \,=\, 1 \,+\, k \,/\,
\alpha^2 \;\; , \end{equation}

\noindent with \,$k \geq 0$\, an arbitrary constant.  For \,$k > 0$,
however, this form of the function \,$f$\, will not be very useful in
spacetimes with large curvatures.  The reason for this is easy to see.
Even thought the condition will prevent the formation of shocks, it
implies an evolution equation for the lapse of the form:
\begin{equation} \partial_t \, \alpha \,=\, -\, \left( \alpha^2 \,+\, k
\, \right) \; {\rm tr} \, K \;\; .  \end{equation}

\noindent Clearly, in a region where the lapse has collapsed to a very
small value we will have:  \begin{equation} \partial_t \, \alpha
\,\simeq\, -\, k \; {\rm tr} \, K \;\; .  \end{equation}

\noindent  If \,${\rm tr}\, K > 0$, there is nothing to prevent the
lapse from becoming negative (this can in fact happen very easily in
black hole simulations).  We are then led to the conclusion that the
only value of \,$f$\, that will prevent the first family of shocks from
developing without carrying the risk of leading to a negative lapse is
\,$f = 1$, {\em i.e} harmonic slicing.

The second family of shocks, on the other hand, is independent of the
form of \,$f$\, and arises even for harmonic slicing.  This is a very
unexpected result.  After all, this is precisely the slicing used to
prove the theorems of existence and uniqueness of solutions in General
Relativity~\cite{Choquet62,HawkingEllis73,Wald84}.  Since at a shock
the differential equations break down, one would expect the theorems to
forbid such solutions.  We must remember, however, that these theorems
can only be proved {\em locally},  they can not therefore rule out a
shock that develops after a {\em finite time}.

It must be stressed that the violation of indirect linear degeneracy is
not a sufficient condition for the development of shocks.  The choice of
initial data will have a crucial effect in whether or not shocks
actually develop.  In particular, since shocks
are a consequence of transport, they should only develop when we have
wave propagation, either in the form of real gravitational waves, or in
the form of pure gauge waves as was shown in the examples presented
here.  Of course, in the simple cases considered in this paper one can
easily find initial data that does not produce shocks:  for the flat
spacetimes one can just take a flat initial slice, while for a black
hole spacetime we can start from an unperturbed Schwarzschild slice.
In the more general case, however, it might be difficult to find such
benign initial data, or even to prove that it exists at all.

One more important point should be made here.  Since the shocks that I
have found arise in the case of a zero shift vector, they must
necessarily indicate a break down of the slicing condition.  That is,
the shocks represent places where the spatial hyper-surfaces become
non-smooth.  Since the presence of a shift vector can not alter the
geometry of these hyper-surfaces, the shocks found here must appear for
{\em any shift condition}.  A given shift might eliminate the
discontinuities in some components of the spatial metric, but it can
not eliminate the shocks completely:  at least some of the dynamical
quantities will remain non-smooth for all possible shift choices.

\vspace{5mm}

Although in this paper I have concentrated in the BM hyperbolic
formalism, the mathematical tools developed can easily be applied to
any other hyperbolic formalism of General Relativity.  One should
expect the phenomena of coordinate shocks to also arise in any such
formalism.  In fact, since all formalisms must have the same physical
solutions, the results of this paper imply that in {\em any
formalism}\/ the use of a harmonic slicing will generate shocks for
at least some initial conditions.

Clearly, the search for gauge conditions and/or restrictions on the
initial data that can prevent the development of coordinate shocks
is a problem that must be addressed if hyperbolic formalisms are to
become an important tool in the study of both theoretical and numerical
relativity.

\section{AKNOWLEDGEMENTS}

The author wishes to thank Gabrielle~D.~Allen, Carles~Bona, Joan~Mass\'{o}
and Bernard~F.~Schutz for many useful discussions and comments.



\pagebreak

\end{document}